\documentclass[pra,bibnotes,twocolumn]{revtex4}
\usepackage{graphicx}
\begin{document}
\draft
\def\ds{\displaystyle}
\title{Fast Frictionless Expansion of an Optical Lattice}
\author{C. Yuce }
\address{ Physics Department, Anadolu University,
 Eskisehir, Turkey}
\email{cyuce@anadolu.edu.tr}
\date{\today}
\begin{abstract}
We investigate fast frictionless expansion of an optical lattice
with dynamically variable spacing (accordion lattice). We design
an expansion trajectory that yields a final state identical to the
initial state up to an irrelevant phase factor. We discuss the
effect of additional force and nonlinear interaction on the fast
frictionless expansion.
\end{abstract}
\maketitle
\section{Introduction}

In an adiabatic process, an eigenstate of a time dependent
Hamiltonian follows its instantaneous eigenstate. However,
adiabatic processes typically need timescales longer than
lifetimes or coherence times of the system, therefore hindering
their applicability. This has motivated researchers to find a way
to speed up the process to reach the same instantaneous adiabatic
eigenstate \cite{bangbang}. Recently, a new technique has been
introduced
\cite{muga1,muga2,muga3,muga4,muga5,muga01,muga03,muga04,yeni6,deneyfast,yeni10,yeni8,yeni9,yeni11,yeni12,yeni13,campo,onofrio,muga02,naka,salamon,berry,inverted,hof,morschson}
and experimentally realized with ultracold $^{87}Rb$ atoms in a
harmonic trap \cite{deneyfast,yeni10}. The technique is based on
engineering the time dependent parameters of the Hamiltonian. The
fast schemes to get the same final state as the one after
adiabatic process is called shortcut to adiabaticity \cite{muga5}
or superadiabatic \cite{morschson}. It is known that fast schemes
are not adiabatic in between initial and final times. In the
experiment \cite{yeni10}, the trap is decompressed by a factor of
$15$ within 35 $ms$, which is much less than typical adiabatic
process of a few seconds. Choi et al. discussed that optimal
cooling of an atomic species may be obtained by means of
sympathetic cooling with another species whose trapping frequency
is dynamically changed to maintain constancy of the
Lewis-Riesenfeld adiabatic invariant \cite{onofrio}. Recently,
nearly perfect fidelity for a two-level quantum system was
achieved experimentally \cite{morschson}. The fast transitionless
mechanism has also been extended to the Gross-Pitaevskii (GP)
equation \cite{muga02}. Most of the investigations so far have
been based on ultracold atoms in a
harmonic trap.\\
In this paper, we will explore the possibility of fast
frictionless expansion for an optical lattice.  We will show that
a correct choice of expansion trajectory allows us to get a final
adiabatic state in a non-adiabatic way. Our method of obtaining
fast adiabatic transformation is to use a lattice with dynamically
variable spacing (accordion lattice)
\cite{accordion3,accordion4,accordion2,accordion1}. A continuous
change of the lattice periodicity from 0.96 $\mu$m to 11.2 $\mu$m
in one second was achieved in an early experiment with an optical
lattice \cite{accordion4}. Recent experiment with ultracold atoms
of $^{87}$Rb trapped in a two-dimensional optical lattice
demonstrated the variation of the spacing of the lattice in two
directions from $2.2 {\mu}m$ to $5.5 {\mu}m$ in a few milliseconds
\cite{accordion1} using dual axis acousto-optic deflectors. Here,
we will give the conditions for fast frictionless expansion for an
accordion lattice. Such accordion lattices are useful since the
final lattice spacing can be made large enough to be resolved in
an experiment. Note that, in principle, the lattice spacing can be
extended to any size smaller than the laser beam waist. This
allows imaging of the atoms at individual sites.

\section{Accordion Lattice}

\begin{figure}
\centering
\includegraphics[width=6cm]{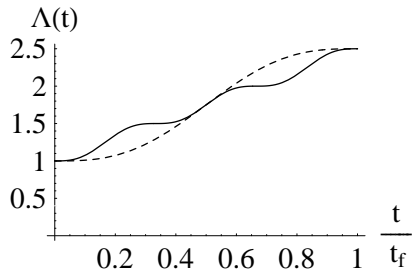}
\caption{$\ds{\Lambda(t)}$ versus $\ds{t/t_f}$ for $\ds{c=2.5}$.
The dashed (solid) curve is for $\ds{\Lambda_1(t)}$
($\ds{\Lambda_2(t)}$ with $\ds{n=3}$). } \label{uzc1}
\end{figure}

\begin{figure}
\centering
\includegraphics[width=6cm]{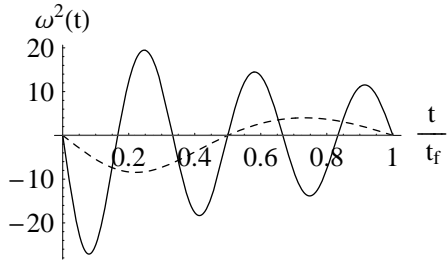}
\caption{$\ds{\omega^2(t)}$ versus $\ds{t/t_f}$ for $\ds{c=2.5}$.
The dashed (solid) curve is for $\ds{\Lambda_1(t)}$
($\ds{\Lambda_2(t)}$ with $\ds{n=3}$).  } \label{uzc2}
\end{figure}

\begin{figure}
\centering
\includegraphics[width=6cm]{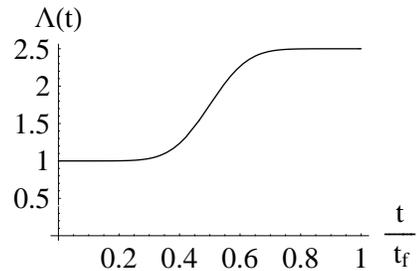}
\caption{ $\ds{\Lambda_3(t)}$ versus $\ds{t/t_f}$. The lattice is
expanded by a factor of $2.5$ from $2.2 {\mu}m$ to $5.5 {\mu}m$ in
a few milliseconds. } \label{uzdeney2}
\end{figure}

\begin{figure}
\centering
\includegraphics[width=6cm]{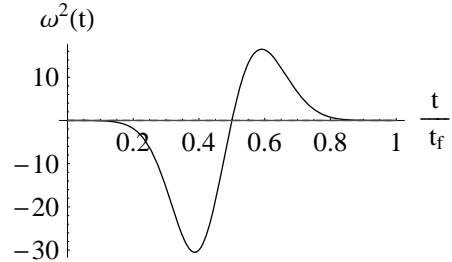}
\caption{  $\ds{\omega^2(t)}$ versus $\ds{t/t_f}$ for
$\ds{\Lambda_3(t)}$} \label{uzdeney1}
\end{figure}

In contrast to electrons in a crystal lattice, optical lattices
allow us to change parameters, such as potential depth and lattice
spacing. The potential depth of the optical lattice can be
modulated by changing the power of the laser. The lattice spacing
can be changed by changing the wavelength of the laser or by
changing the relative angle between the two laser beams
\cite{accordion3,accordion4,accordion2,accordion1}. Let us
consider ultracold atoms in an optical lattice with variable
spacing and potential depth. We assume that the transverse motion
of the atoms is frozen (i.e., we are dealing with a 1-D problem).
The description of a condensate in 1-D is based on the Hamiltonian
\begin{equation}\label{3body}
H=\frac{p^2}{2m} +V(t)\cos{\left(2k_L\frac{x}{\Lambda(t)}\right)}
+\frac{m \omega^2(t)}{2}x^2 ~.
\end{equation}
where $\ds{m}$ is the atomic mass, $\ds{V(t)}$ is the lattice
depth, $\ds{k_L}$ is the optical lattice wave number,
$\ds{\omega(t)}$ is the time dependent angular frequency and
$\ds{\Lambda(t)}$ is the scale parameter describing the expansion
of the lattice. The initial and final lattice scalings are given
by
\begin{equation}\label{sart2}
{\Lambda}(t=0)=1~,~~{\Lambda}(t=t_f)=c~,
\end{equation}
where the constant $c$ is the ratio of the final spacing to the
initial spacing.\\
Initially only the periodic potential is present in the system.
When the system starts to expand by changing the lattice spacing,
the parabolic potential is also added. The combined presence of
the periodic and parabolic potentials enables us to get fast
frictionless expansion. Finally, the parabolic potential is turned
off at $t=t_f$ such that
\begin{equation}\label{sart0}
\omega(t=0)=\omega(t=t_f)=0~.
\end{equation}
The system is initially in the stationary state. To find the time
evolution of the state for the Hamiltonian (\ref{3body}), we
introduce a transformation on the wave function as
\begin{eqnarray}\label{trans1}
\Psi(x,t)=\exp\left(-igx^2/2+\frac{\hbar}{2m}\int g
dt\right)~\Phi(x,t)~,
\end{eqnarray}
with a subsequent transformation on the coordinate
\begin{eqnarray}\label{trans2}
x=z\Lambda(t)~.
\end{eqnarray}
where $g(t)$ is a time dependent function to be determined later.
Indeed, the scale transformation on the coordinate accounts for
the expansion of the condensate. Now let us substitute these
transformations into the corresponding Schrodinger equation for
the Hamiltonian (\ref{3body}). Note that due to the scaling of
coordinate, the time derivative operator transforms as
$\ds{\partial/\partial t \rightarrow
\partial/\partial t-(\dot{\Lambda}/\Lambda) z~ \partial/\partial z}$,
where dot denotes time derivation. Choosing
\begin{eqnarray}\label{gfactor}
g(t)=-\frac{m}{\hbar}\frac{\dot{\Lambda}}{\Lambda}~,
\end{eqnarray}
yields
\begin{eqnarray}\label{3body2}
i\hbar\frac{\partial \Psi}{\partial
t}=-\frac{\hbar^2}{2m\Lambda^2}\frac{\partial^2 \Psi}{\partial
z^2}+V \cos(2k_Lz)\Phi+\frac{m}{2}\Omega^2 z^2 \Phi
\end{eqnarray}
where $\Omega^2$ is given by
\begin{eqnarray}\label{Omega}
\Omega^2(t)=\Lambda^2\omega^2+\Lambda\frac{\partial^2{\Lambda}}{{\partial}t^2}~.
\end{eqnarray}
To sum up, the effects of lattice spacing are equivalent to the
effective time dependent mass and the effective parabolic
potential in the stationary frame. The first term in the right
hand side of (\ref{Omega}) is due to the external one while the
second one is induced due to the variation of the lattice spacing.
Indeed, the last formula explains the difficulties of keeping
atoms trapped in an experiment with an accordion lattice
\cite{accordion2,accordion1}. In the absence of external harmonic
trap, $\omega^2=0$, the system is left with the induced parabolic
potential that depends on the acceleration,
$\ds{\frac{\partial^2{\Lambda}}{{\partial}t^2}}$. Since the
expansion starts at $t=0$ and stops at $t=t_f$, the acceleration
takes both positive and negative values in the interval $0<t<t_f$.
Hence the repulsive potential with an imaginary frequency,
$\Omega^2<0$, has the effect of pushing the atoms out of the
condensate. To tackle the problem, the lattice depth was ramped
from $ h\times1.5 $ $Hz$ to $ h\times37 $ $Hz$ in $5~ms$ in the
experiment \cite{accordion1}. Here we instead suggest decreasing
the lattice depth. The first reason for this is that the fast
frictionless expansion can be achieved by engineering the
potential depth as
\begin{equation}\label{V0}
V(t)= \frac{V_0}{\Lambda^2(t)}~.
\end{equation}
Hence the potential depth decreases as the system expands.
Secondly, we demand that no effective harmonic potential acts on
the system, $\ds{\Omega^2=0}$. In other words, we choose
$\ds{\omega^2 (t)}$ in such a way that it cancels the effective
parabolic potential so that there is only periodic potential in
the system from $\ds{t=0}$ to $\ds{t=t_f}$. Hence, we choose
\begin{eqnarray}\label{Omegaesitsýfýr}
\omega^2(t)=-\Lambda^{-1}\frac{\partial^2{\Lambda}}{{\partial}t^2}~.
\end{eqnarray}
This choice of external frequency guarantees that ultracold atoms
stay inside the optical lattice. \\
The conditions (\ref{V0},\ref{Omegaesitsýfýr}) reduce the equation
(\ref{3body2}) to the following one
\begin{eqnarray}\label{3body3}
i\hbar\frac{\partial
\Phi}{\partial\tau}=-\frac{\hbar^2}{2m}\frac{\partial^2
\Phi}{\partial z^2}+V_0 \cos(2k_L z)\Phi
\end{eqnarray}
In the last step, we made another transformation on time
$\ds{\tau=\int_0^t \Lambda^{-2}(t^{\prime})~ dt^{\prime}}$.\\
It is interesting to observe that the Schrodinger equation for the
Hamiltonian (\ref{3body}) can be transformed into the another
Schrodinger equation (\ref{3body3}) with the constant potential
depth. We can view (\ref{3body3}) as describing another condensate
with a different time parameter.\\
We find a new Hamiltonian $\ds{H^{\prime}=\frac{p^2}{2m}+V_0
\cos(2k_L z)}$ for the system described by the Hamiltonian
(\ref{3body}). The final state of $H^{\prime}$ at $t_f$ can be
made identical up to a global phase factor to the final state of
the adiabatic evolution with $H$ if we impose some conditions on
the scale parameter $\ds{\Lambda(t)}$. By virtue of
(\ref{trans1},\ref{gfactor}), we conclude that they are given by
\begin{equation}\label{sart1}
\dot{\Lambda}(0)=\dot{\Lambda}(t_f)=0~,~~\ddot{\Lambda}(0)=\ddot{\Lambda}(t_f)=0~,
\end{equation}
where dot is the time derivation with respect to the parameter
$t$. The acceleration, $\ddot{\Lambda}$, is set to zero at initial
and final times since the initial and final external frequencies
are assumed to be zero (\ref{sart0},\ref{Omegaesitsýfýr}). These
conditions together with the conditions (\ref{sart2})
guarantee the fast transitionless evolution. \\
Here we will mainly be interested in three different solutions for
$\ds{\Lambda(t)}$.  The first one satisfying the relations
(\ref{sart2},\ref{sart1}) is the polynomial solution
\begin{eqnarray}\label{cozum}
\Lambda_1(t)=(c-1)\left(\frac{1}{c-1}+10 \frac{t^3}{t_f^3}-15
\frac{t^4}{t_f^4}+6 \frac{t^5}{t_f^5}\right)~,
\end{eqnarray}
and the second one is given by
\begin{eqnarray}\label{cozum2}
\Lambda_2(t)=(c-1)\left(\frac{1}{c-1}+
\frac{t}{t_f}-\frac{1}{2n\pi} \sin(\frac{2n{\pi}t}{t_f})\right)~.
\end{eqnarray}
where $n$ is a positive integer. For these trajectories, the
external frequency can be calculated using (\ref{Omegaesitsýfýr}).
In Fig-{\ref{uzc1}} and Fig-{\ref{uzc2}}, we plot $\ds{\Lambda}$
and corresponding $\ds{\omega^2}$. The dashed (solid) curve is for
$\ds{\Lambda_1}$ ($\ds{\Lambda_2}$ with $n=3$). We assume the
lattice is expanded by a factor of $2.5$. As  expected, the sign
of $\omega^2(t)$ changes in time. There are some differences in
implementing fast frictionless processes in harmonic traps and
optical lattices. For a harmonically trapped system, we need the
harmonic potential to achieve the expansion of the system.
However, expansion for an optical lattice can be achieved either
by changing the wavelength of the laser or the relative angle
between the two laser beams. The external parabolic potential is
necessary for an optical lattice to cancel the induced parabolic
potential, which pushes the atoms out of the condensate since it
becomes expulsive for certain time intervals. \\
One interesting trajectory was already realized in the experiment
\cite{accordion1}, where the lattice was expanded by a factor of
$2.5$ in a few milliseconds
\begin{eqnarray}\label{cozum3}
\Lambda_3(t)=  1+0.75~\left(\mathrm{erf}\{7.14
(\frac{t}{t_f}-0.5)\} +1\right)  ~.
\end{eqnarray}
where $\mathrm{erf}$ is the Gauss error function.
$\ds{\Lambda_3(t)}$ satisfies (\ref{sart1}) to a very good
approximation. Fig-{\ref{uzdeney2}} and Fig-{\ref{uzdeney1}} plot
$\ds{\Lambda_3(t)}$ and corresponding $\ds{\omega^2(t)}$. As can
be seen, $\ds{\omega^2}$ is negative between $0<t<t_f/2$, while it
is positive otherwise. We propose to repeat the experiment [28]
with the same trajectory but a different potential depth and a
different frequency. If they change according to
(\ref{V0},\ref{Omegaesitsýfýr}), then fast frictionless expansion
can
be realized.\\
So far, we have studied 1-D optical lattice with variable spacing.
The generalization to higher dimensions is straightforward. In
this section, we will discuss the effects of additional force,
harmonic potential and scattering length on the fast frictionless
expansion. Let us first consider the presence of an additional
force, $F(t)$, in the Hamiltonian (\ref{3body}). The force can be
a constant gravitational force or a periodically changing one. The
latter is experimentally realized by generating periodical
frequency shift between the two laser beams \cite{dynloc}. The
dynamical responses of the particle to the constant force and
periodically changing force are the well-known phenomena of Bloch
oscillations \cite{bloch,bloch2,bloch3} and dynamical localization
\cite{dunlap}, respectively. The additional force, $F(t)$
transforms as $F(\tau)\Lambda^3(\tau)$ in the equation
(\ref{3body3}). Hence, the picture for Bloch oscillation and
dynamical localization change a lot for the trajectories
(\ref{cozum},\ref{cozum2},\ref{cozum3}). The fast frictionless
expansion without destroying the Bloch oscillation and dynamical
localization would be possible if the force
transforms as $F/\Lambda^3$.\\
Suppose now that the system is initially prepared in the combined
presence of the periodic plus parabolic potentials. Hence, the
initial and final frequencies are given by
$\ds{\omega(0)=\omega(t_f)=\omega_0}$. In this case, the condition
(\ref{sart1}) is still valid but (\ref{Omegaesitsýfýr}) is not.
The fast frictionless expansion for this case is possible if
$\ds{\Omega^2}$ is given by $\ds{\Omega^2=\omega_0^2/\Lambda^2}$
in the equation (\ref{3body2}). The specific external frequency,
$\omega(t)$, satisfying this condition can be found easily using
the equation (\ref{Omega}). Let us now discuss dipole oscillation.
In the experiment for 3-D BEC \cite{deney1}, the trap in the
presence of the optical lattice was suddenly shifted. It was
observed that the subsequent oscillations of the BEC are undamped
if the initial displacement is small, but become dissipative if
the initial displacement exceeds a critical value. Fertig et al.
\cite{fertig} induced dipole oscillations of 1-D Bose gases for
different values of the axial lattice depth. They observed a
transition from underdamping to overdamping with increasing
lattice depth. Due to strong quantum fluctuations, mean-field
theories fail for the description of the system. Consider an
experiment like the ones in \cite{deney1,deney2} to study dipole
oscillation for an accordion lattice. The amplitude of the small
dipole oscillation increases with $\Lambda$ without leading
instability since the lattice is expanding (\ref{trans2}). The fast frictionless expansion preserves dipole oscillation.\\
Let us now discuss the extension of the fast frictionless
expansion to GP equation. If the scattering length is changed
according to $\ds{a_s(t)=a_s(0)/\Lambda^2}$ via Feshbach
resonance, then adiabatic final state can be obtained
non-adiabatically. As a result, we say that the fast frictionless
expansion is not limited to
noninteracting case.\\
In conclusion, we have shown that the non-adiabatic expansion of
an optical lattice in few milliseconds may lead to adiabatic final
states provided that the expansion trajectory satisfies
(\ref{sart1}) and the external parabolic potential is applied to
the system. This prediction can be tested through experiments
similar to the one described in \cite{accordion1}. Since the
non-adiabatic protocol preserves the initial state after
expansion, it allows straightforward imaging of the optical
lattice. We have also discussed the effect of additional forces
and harmonic trapping. Finally, we argue that fast frictionless
expansion is possible even in the presence of nonlinear
interaction provided that the scattering length is decreased
proportionally to $\Lambda^2$ via Feshbach resonance.


\begin{thebibliography}{00}
\bibitem{bangbang} P. Salamon, K. H. Hoffmann, Y. Rezek, and R. Kosloff, Phys. Chem. Chem. Phys. {\bf11}, 1027 (2009).
\bibitem{muga1} Xi Chen, E. Torrontegui, Dionisis Stefanatos, Jr-Shin Li, J. G. Muga, Phys. Rev. A {\bf84}, 043415 (2011).
\bibitem{muga2} S. Ib\'{a}\~{n}ez, S. Martí nez-Garaot, Xi Chen, E. Torrontegui, J. G. Muga, Phys. Rev. A {\bf84}, 023415 (2011).
\bibitem{muga3} Xi Chen, E. Torrontegui, J. G. Muga, Phys. Rev. A {\bf83}, 062116 (2011).
\bibitem{muga4} E. Torrontegui, S. Ibanez, Xi Chen, A. Ruschhaupt, D. Guery-Odelin, J. G. Muga, Phys. Rev. A {\bf83}, 013415 (2011).
\bibitem{muga5} Xi Chen, I. Lizuain, A. Ruschhaupt, D. Guery-Odelin, J. G. Muga, Phys. Rev. Lett. {\bf105}, 123003 (2010).
\bibitem{muga01}X. Chen and J. G. Muga, Phys. Rev. A {\bf82}, 053403 (2010).
\bibitem{muga03} X. Chen, A. Ruschhaupt, S. Schmidt, A. del Campo, D. Guery-Odelin, and J. G. Muga, Phys. Rev. Lett. {\bf104}, 063002 (2010).
\bibitem{muga04} J. G. Muga, X. Chen, S. Ibanez, I. Lizuain, and A. Ruschhaupt, J. Phys. B {\bf43}, 085509 (2010).
\bibitem{yeni6} S. Masuda and K. Nakamura, Proc. R. Soc. A {\bf466}, 1135 (2010).
\bibitem{deneyfast} J. F. Schaff, X. L. Song, P. Vignolo, and G. Labeyrie, Phys. Rev. A {\bf82}, 033430 (2010).
\bibitem{yeni10} J. F. Schaff, X. L. Song, P. Capuzzi, P. Vignolo, and G. Labeyrie, Europhys. Lett. {\bf93}, 23001 (2011).
\bibitem{yeni8} D. Stefanatos, J. Ruths, and Jr-Shin Li, Phys.Rev. A {\bf82}, 063422 (2010).
\bibitem{yeni9} S. Masuda and K. Nakamura, Physical Review A {\bf84}, 043434 (2011).
\bibitem{yeni11} J. F. Schaff, P. Capuzzi, G. Labeyrie and P. Vignolo, New J. Phys. {\bf13} 113017 (2011).
\bibitem{yeni12} B. Andresen, K. H. Hoffmann, J. Nulton, A. Tsirlin, and P. Salamon, Eur. J. Phys. {\bf32}, 827 (2011).
\bibitem{yeni13} Y. Li, L.-A. Wu, and Z.-D. Wang, Phys. Rev. A {\bf83}, 043804 (2011).
\bibitem{campo} A. del Campo, Phys. Rev. A {\bf84}, 031606(R) (2011).
\bibitem{onofrio} S. Choi, R. Onofrio, and B. Sundaram, Phys. Rev. A {\bf84} 051601(R) (2011).
\bibitem{naka} M. Fasihi, Y. Wan, M. Nakahara, J. Phys. Soc. Jpn. {\bf81}, 024007 (2012).
\bibitem{salamon} B. Andresen, K. H. Hoffmann, J. Nulton, A. Tsirlin and P. Salamon, Eur. J. Phys. {\bf32} 827 (2011).
\bibitem{berry} M. V. Berry, J. Phys. A {\bf42}, 365303 (2009).
\bibitem{inverted} C Yuce, A Kilic and A Coruh, Phys. Scr. {\bf 74} 114 (2006).
\bibitem{hof} K. H. Hoffmann, et. al, EPL {\bf 96} 60015 (2011).
\bibitem{morschson} Mark G. Bason, et. al, Nature Phys. {\bf 8}, 147 (2012).
\bibitem{muga02} J. G. Muga, X. Chen, A. Ruschhaupt, and D. Gu\'{e}ry- Odelin, J. Phys. B {\bf42}, 241001 (2009).
\bibitem{accordion3} Leonardo Fallani, Chiara Fort, Jessica Lye, Massimo Inguscio, Opt. Express {\bf13}, 4303 (2005).
\bibitem{accordion4} T. C. Li, H. Kelkar, D. Medellin, M. G. Raizen, Opt. Express {\bf16}, 5465 (2008).
\bibitem{accordion2} R. A. Williams, J. D. Pillet, S. Al-Assam, B. Fletcher, M. Shotter, C. J. Foot, Optics Express, {\bf21}, 16977 (2008).
\bibitem{accordion1}S. Al-Assam, R. A. Williams, C. J. Foot, Phys. Rev. A {\bf82}, 021604(R) (2010).
\bibitem{dynloc} H. Lignier, C. Sias, D. Ciampini, Y. Singh, A. Zenesini, O. Morsch, E. Arimondo, Phys. Rev. Lett. {\bf99}, 220403 (2007).
\bibitem{bloch} M. Ben Dahan, E. Peik, J. Reichel, Y. Castin, and C. Salomon, Phys. Rev. Lett. {\bf 76}, 4508 (1996).
\bibitem{bloch2} O. Morsch, J. H. Muller, M. Cristiani, D. Ciampini, and E. Arimondo, Phys. Rev. Lett. {\bf 87}, 140402 (2001).
\bibitem{bloch3} B. P. Anderson and M. Kasevich, Science {\bf 282}, 1686 (1998).
\bibitem{dunlap} D. H. Dunlap and V. M. Kenkre, Phys. Rev. B {\bf34}, 3625 (1986).
\bibitem{deney1} S. Burger, F. S. Cataliotti, C. Fort, F. Minardi, M. Inguscio, M. L. Chiofalo, M. P. Tosi, Phys. Rev. Lett. {\bf86}, 4447 (2001).
\bibitem{deney2} A. Smerzi, A. Trombettoni, P.G. Kevrekidis, A.R. Bishop, Phys. Rev. Lett. {\bf89}, 170402 (2002).
\bibitem{fertig} C. D. Fertig et al., Phys. Rev. Lett. {\bf94}, 120403 (2005).
\end{thebibliography}
\end{document}